\documentclass[aps,pra,twocolumn,amssymb, amsfonts,amsmath,showpacs, superscriptaddress]{revtex4}
 \usepackage{graphicx}
  \usepackage{amssymb}
  \begin{document}
\title{Cosmogenesis and Collapse}
\author{Philip Pearle}
\affiliation{Emeritus, Department of Physics, Hamilton College, Clinton, NY  13323}
\email{ppearle@hamilton.edu}
\begin{abstract}
 {Some possible  benefits of dynamical collapse for a quantum theory of cosmogenesis are discussed.   These are a possible long wait before creation begins, creation of energy and space, and choice of a particular universe out of a superposition.}  
 \end{abstract}
\maketitle

\section{Introduction}\label{sec1} 

Suppose one believes that the onset of our universe should be described by quantum theory. Quantum evolution famously produces a state vector which is a superposition of possibilities, but our particular universe is just one of those possibilities.  How was the choice made?  A dynamical collapse theory can help with this. 

 Such a theory as CSL (Continuous Spontaneous Localization) \cite{CSL,reviews}, constructed to describe wave function collapse within the present universe, works quite well.  The state vector evolves continuously,  governed by a Hamilton whose Hermitian part is the usual Hamiltonian, and whose anti-hermitian part depends upon a randomly fluctuating field. When the hermitian Hamiltonian evolves the state vector into a superposition of macroscopically different states,  the anti-hermitian Hamiltonian causes a rapid evolution to one of them. 

A dynamical collapse theory posits that the state vector corresponds to the real state of the world, and provides a mechanism to accomplish this. If one believes that nature uses the collapse mechanism to choose the outcome of an everyday event, it is natural to suppose that a similar mechanism was at work for the Big Event. 

Suppose  the pre-universe is modeled by a state $|0\rangle$ which represents no space.  Further, suppose that time exists. (An eventual hope might be to have time come out of such a theory, but  crawling comes  before walking.)  

A Hamiltonian  must be chosen to evolve the state $|0\rangle$ into our early universe.  
One might suppose this is a relatively rare event, i.e.,  the universe has a low probability/sec of being created out of the pre-universe.  Dynamical collapse can help with this.  

In standard quantum theory with the instantaneous collapse postulate, there is a phenomenon  called the ``Zeno" or ``watched pot" effect.  Suppose a system starts in state $|0\rangle$ and evolves under Hamiltonian $H$ for a short time $\Delta t$, followed by instantaneous collapse.   Suppose this occurs repeatedly over the time interval $T$. It is straightforward to show that the probability the system remains in state $|0\rangle$ is $\exp-[(\Delta H)^{2}T\Delta t+\hbox{o}(\Delta t)^{2}]$, 
where $(\Delta H)^{2}\equiv  \langle 0|H^{2}|0\rangle - \langle 0|H|0\rangle^{2}$.  Thus, as $\Delta t$ gets smaller, the probability that $|0\rangle$ evolves is reduced.  A comparable behavior occurs for continuous collapse. When the collapse rate parameter $\lambda$ increases sufficiently, the probability that the initial state vector evolves is reduced.  

Not only space needs to be created, but also the energy which fills that space.  Dynamical collapse can help with this.  

In usual quantum theory, energy is conserved.  However, that need not be the case in a dynamical collapse theory.  For example, in CSL, particle wave functions are narrowed by the collapse process, which means that their energy grows\cite{pearlesquiresetc.}.  

In this paper, a simple cosmogenesis model to illustrate these features is presented. Its mathematics is that of a spin 1/2  caused to rotate with angular frequency $\omega$ by a constant $y$-directed magnetic field, and that of a perturbed harmonic oscillator, both subject to collapse dynamics. 

 The spin 1/2 acts like a gatekeeper to block (state  $|\uparrow\rangle$) or allow (state $|\downarrow\rangle$) the harmonic oscillator to operate.  The spin starts in the block state but is prevented from rotating with frequency $\omega$  because of ``watched pot" behavior:  a sufficiently large value of the collapse rate parameter $\lambda$ can make  the probability/sec of a transition to the allow state arbitrarily small.  
 
The harmonic oscillator starts in the ground state $|0\rangle$ representing no space volume.  The states  $|n\rangle$ represent a universe of  $n$ cells, with energy $\epsilon n$ and volume $vn$. To be concrete, we shall  suppose that  $\epsilon$=Planck energy.   $v$=Planck volume is a possible choice, or $v$ may be an increasing function of time.  One may consider that the energy generated is the dark energy alone, or all the energy: the nature of the energy and the size of $v$ do not play a role in the dynamics (but see section \ref{secV}). 

Due to a perturbation,  the oscillator would just oscillate between $|0\rangle$  and slightly higher $|n\rangle$ states.  (Two kinds of perturbation shall be considered, an added term proportional to position in section \ref{secIII} and to squared position in \ref{secIV}.)  However,  because of the attendant collapse (rate parameter $\lambda'$), the universe is generated: the state vector evolves states of  increasing values of $n$. With the initial state $|\uparrow>|0\rangle$,  the (unnormalized) state vector at time $t$ is taken to be

\begin{eqnarray}\label{1}
|\Psi, t\rangle_{w,w'}&=&{\cal T}e^{-\int_{0}^{t}dt'\{\frac{i}{2}\omega\sigma_{2} +\frac{1}{4\lambda}[w(t')-2\lambda\sigma_{3}]^{2}\}} \nonumber\\
&&\negthinspace \negthinspace\negthinspace\negthinspace\negthinspace\negthinspace
\negthinspace \negthinspace\negthinspace\negthinspace\negthinspace\negthinspace
\negthinspace \negthinspace\negthinspace\negthinspace\negthinspace\negthinspace
\negthinspace \negthinspace\negthinspace\negthinspace\negthinspace\negthinspace
\negthinspace \negthinspace\negthinspace\negthinspace\negthinspace\negthinspace
\negthinspace \negthinspace\negthinspace\negthinspace\negthinspace\negthinspace
\negthinspace \negthinspace\negthinspace\negthinspace\negthinspace\negthinspace
\cdot e^{-\int_{0}^{t}dt'\{i[\epsilon a^{\dagger}a+g(a^{\dagger}+a)^{s}]Q+\frac{1}{4\lambda'} [w'(t')-2\lambda' a^{\dagger}a]^{2}\}}|\uparrow>|0\rangle.
\end{eqnarray}

 In Eq.(\ref{1}), ${\cal T}$ is the time-ordering operator.  If there were no collapse, the spin would rotate  clockwise in the $x-z$ plane. $Q\equiv|\downarrow\rangle\langle\downarrow|$ is the projection operator on the allow state. $g$ is a coupling constant:  we shall consider the cases of $s=1$ (an interaction which displaces the harmonic oscillator) and  $s=2$ (an interaction which changes its spring  constant). 
 
 $w(t')$ and  $w'(t')$ are white noise random functions of time. The probability density for $w$, $w'$ to take on particular values over the time interval $(0\leq t'\leq t)$ is given by the probability rule:
 \begin{equation}\label{2}
P(w(t), w'(t))\sim_{w,w'}
\negthinspace\negthinspace
\langle\Psi,t|\Psi, t\rangle_{w,w'}
\end{equation}
\noindent where the proportionality constant is determined by 
\[
\int_{-\infty}^{\infty}DwDw'P=1
\]
\noindent with $DwDw' \equiv \prod_{i}dw(t_{i})dw'(t_{i})$ ($t_{i+1}-t_{i}\equiv dt$).  

	  According to Eq.(\ref{1}), each different $w(t')$, $w'(t')$ gives rise to a different state vector $|\Psi, t\rangle_{w,w'}$.  The probability of that state vector (of that $w(t')$, $w'(t')$) is given by Eq.(\ref{2}).  
  Eqs.(\ref{1},\ref{2}) completely define the model. It only remains to draw the consequences.  
  
In this paper, we shall only consider the  evolutions of the spin and harmonic oscillator separately, not their joint evolution, in order to keep the discussion simple and to elucidate the features of each.   
 
\section{Gatekeeper/Spin}\label{sec2} 

	A numerical analysis has been given for this problem\cite{IB}.  Here we give an analytical treatment.  
	
	The Fokker-Planck equation for the probability distribution $R(\theta,t)$ of the angle $0\leq\theta\leq 2\pi$ that the spin makes with the $z$-axis is
\begin{equation}\label{3}
\frac{\partial}{\partial t}R(\theta,t)=-\frac{\partial}{\partial \theta}\Big[\frac{\overline{d\theta}}{dt}R(\theta,t)\Big]+\frac{1}{2}\frac{\partial^{2}}{\partial \theta^{2}}\Big[\frac{\overline{(d\theta)^{2}}}{dt}R(\theta,t)\Big],
\end{equation}
\noindent where the average is over all possible trajectories of $\theta$ due to all possible $w(t)$. Appendix \ref{A}  shows how to calculate  the drift $\overline{d\theta}/dt$ and the diffusion $\overline{d\theta^{2}}/dt$ (Eqs.(\ref{A6},\ref{A7})), yielding the Foker-Planck equation: 
\begin{equation}\label{4}
\frac{\partial}{\partial t}R=-\frac{\partial}{\partial \theta}[\omega-2\lambda\sin\theta\cos\theta]R+2\lambda\frac{\partial^{2}}{\partial \theta^{2}}\sin^{2}\theta R. 
\end{equation}

	Qualitatively, we see that the drift's steady clockwise rotation with angular velocity $\omega$ is opposed by the stochastic drift in the first and third quadrants, and aided in the second and fourth  quadrants: the stochastic drift tends to drive the spin to the nearest pole ($\theta=0, \pi$).  The diffusion rate is largest at $z=0$. The diffusion rate approaches 0 at the poles $\sim \theta^{2}, (\pi-\theta)^{2}$. This makes the poles absorbing boundaries\cite{Pearletimetoreduce}: if $\omega=0$, all the trajectories eventually end up at the poles (collapse).  
	
\subsection{$\omega=0$}\label{subIIA}
	
	We begin by considering  uncontaminated collapse behavior ($\omega=0$).  Although Eq.(\ref{4}) is exactly soluble (see \cite{pearle76}, restated in Appendix \ref{C} in the notation of this paper), we prefer to examine its properties in a way that is transferrable to the $\omega\neq0$ case.  
	
	Here is the general context\cite{pearle76}.  Consider an ensemble of normalized state vectors $|\psi, t \rangle =\sum_{k=1}^{N}c_{k}(t)|k\rangle$ which evolve stochastically, but start out identically at $t=0$. Two conditions on $x_{k}(t)\equiv |c_{k}(t)|^{2}$ are sufficient  to obtain collapse behavior obeying the Born rule.  These are the Martingale condition $\overline{x_{k}(t)}=x_{k}(0)$ and the asymptotic correlation condition $\overline{x_{k}(\infty)x_{l}(\infty)}=0$ ($k\neq l$). 
	
	The latter equation is an integral over the product of non-negative terms.  The only way it can be satisfied is if, for every state, every  $x_{k}(\infty)$  but one vanishes.  Therefore, the asymptotic probability density distribution must be	
\[
P(x_{1}, ... x_{n}, \infty)=a_{1}\delta( 1-x_{1})\delta(x_{2}) ... \delta(x_{N})+ ...   
\]
\noindent The Martingale condition then implies $a_{k}=x_{k}(0)$, i.e., the Born rule.

	For our problem, there are two states: $x_{1}(t)=\cos^{2}[\theta(t)/2]$ is the squared amplitude of $|\uparrow\rangle$ and  $x_{2}(t)=\sin^{2}[\theta(t)/2]$ is the squared amplitude of $|\downarrow\rangle$.  We use (\ref{4}) to calculate $\overline {\cos\theta}(t)$, $\overline {\sin^{2}\theta}(t)$.  After integrations by parts, we obtain:
	
\begin{eqnarray}\label{5}
\frac{\partial}{\partial t}\overline{\cos\theta}(t)&\equiv&\frac{\partial}{\partial t}\int_{0}^{2\pi}d\theta\cos\theta R=0\\
\frac{\partial}{\partial t}\overline{\sin^{2}\theta}(t)&\equiv&\frac{\partial}{\partial t}\int_{0}^{2\pi}d\theta\sin^{2}\theta R=-4\lambda\overline{\sin^{4}\theta}(t).\label{6}
\end{eqnarray}

 It follows from Eq.(\ref{5}) that the squared amplitude $(1/2)(1+\overline{\cos\theta})=\overline{\cos^{2}(\theta/2)}$ is constant, so the Martingale condition is  satisfied.  
 
Now,  $\overline{\sin^{2}\theta}(t)$ is non-negative but, according to Eq.(\ref{6}), it keeps decreasing unless $\overline{\sin^{4}\theta}(t)$ vanishes. Therefore,  $\overline{\sin^{4}\theta}(t)$ must vanish at some time 
$t=T$.  But, that is only possible if $R(\theta, T)$ is non-zero only where $\sin^{4}\theta$ is zero, i.e., at $\theta=0, \pi$.  But, this means that  $\overline{\sin^{2}\theta}(T)=0$ too.  Since the average of the product of the up and down amplitudes is $\overline{\sin^{2}[\theta/2]\cos^{2}[\theta/2]}(T)=(1/4)\overline{\sin^{2}\theta}(T)=0$, the asymptotic correlation condition is therefore satisfied. Thus, there is collapse satisfying the Born condition.  
\subsection{$\omega\neq0$}\label{subIIB}	
	It follows from  Eq.(\ref{4}) that
\begin{eqnarray}\label{7}
\frac{\partial}{\partial t}\overline{\cos\theta}(t)&=&-\omega\overline{\sin\theta}(t)\\
\frac{\partial}{\partial t}\overline{\sin\theta}(t)&=&\omega\overline{\cos\theta}(t)-2\lambda\overline{\sin\theta}(t),\label{8}
\end{eqnarray}
\noindent whose solution is
\begin{eqnarray}\label{9}
&&\overline{\cos\theta}(t)=\frac{1}{\lambda_{+}-\lambda_{-}}\Big\{\cos\theta_{0}[\lambda_{+}e^{-\lambda_{-}t}-\lambda_{-}e^{-\lambda_{+}t}]\nonumber\\
&&\qquad\qquad-\omega\sin\theta_{0}[e^{-\lambda_{-}t}-e^{-\lambda_{+}t}]\Big\},\\
&&\lambda_{\pm}\equiv\lambda\pm\sqrt{\lambda^{2}-\omega^{2}}\approx2\lambda, \frac{\omega^{2}}{2\lambda}\quad \hbox{for}\quad\lambda>>\omega.\nonumber
\end{eqnarray}
\noindent where $\theta_{0}$ is the value of $\theta$ at $t=0$. 

	Consider the initial state  $\theta_{0}=0$, spin up. The ensemble average squared amplitude of $|\uparrow\rangle$ is  $\overline{\cos^{2}[\theta/2}](t)=.5[1+\overline{\cos\theta}(t)]$. Thus, according to   Eq.(\ref{9}),  	
	\begin{eqnarray}\label{10}
\overline{\cos^{2}[\theta/2}](t)&=&\frac{1}{2(\lambda_{+}-\lambda_{-})}\Big\{\lambda_{+}[1+e^{-\lambda_{-}t}]-\nonumber\\
&&\qquad\qquad\qquad\quad\lambda_{-}[1+e^{-\lambda_{+}t}]\Big\}\nonumber\\
&\approx&\frac{1}{2}\Big[1+e^{-(\omega^{2}/2\lambda)t}\Big].  
\end{eqnarray}
\noindent Eq.(\ref{10}) shows the "watched pot" behavior mentioned earlier.  For $\omega^{2}/\lambda<<1$, most trajectories remain in the neighborhood of $\theta=0$, i.e., there is only a small probability for a transition from $|\uparrow\rangle$ to $|\downarrow\rangle$.

	Although it is an irrelevant feature for our application of this model,  we close by remarking that, eventually, an equilibrium distribution of spins is reached, which is independent of the original state.  When $R(\theta, t)$ is taken to be independent of $t$, Eq.(\ref{4}) provides the first order differential equation
\begin{equation}\label{11}
-[\omega-2\lambda\sin\theta\cos\theta]R_{eq}(\theta)+2\lambda\frac{\partial}{\partial \theta}\sin^{2}\theta R_{eq}(\theta)=-C,
\end{equation}
 \noindent whose solution is
 \begin{equation}\label{12}
R_{eq}(\theta)=C\frac{1}{|\sin\theta|^{3}} \int_{0}^{\infty}dz\frac{e^{-\frac{\omega}{2\lambda}z}}{[1+(z-\cot\theta)^{2}]^{3/2}}
\end{equation}
According to Eq.(\ref{12}), the equilibrium probability distribution is largest at $\theta=0+, \pi+$, and falls off as $\theta$ increases from 0 to $\pi-$ and from $\pi+$ to $2\pi-$, with a step-up discontinuity crossing the poles in the clockwise direction. 

\section {Spacegenerator/Oscillator I}\label{secIII}

	The state vector evolution we consider here is, from Eq.(\ref{1}), 
\begin{eqnarray}\label{13}
|\Psi, t\rangle_{w}&=&{\cal T}e^{-\int_{0}^{t}dt'\{iH+\frac{1}{4\lambda} [w(t')-2\lambda N]^{2}\}}|0\rangle.
\end{eqnarray}
\noindent where $H=\epsilon N+g(a^{\dagger}+a)$ and $N\equiv a^{\dagger}a$.  The density matrix follows from Eqs.(\ref{1},\ref{2}): 
\begin{eqnarray}\label{14}
\rho(t)&=&\int Dw\medspace_{w}\langle\Psi,t|\Psi,t\rangle_{w} \frac{|\Psi, t\rangle_{w}\negthinspace\negthinspace\negthinspace\negthinspace\negthinspace\quad_{w}\langle\Psi, t|}{\thinspace_{w}\langle\Psi,t|\Psi,t\rangle_{w}}\nonumber\\
&=&{\cal T}e^{-\int_{0}^{t}dt'\{i[H_{L}-H_{R}]+\frac{\lambda}{2}([N_{L}-N_{R}]^{2}\}}|0\rangle\langle 0|,
\end{eqnarray}
\noindent where an operator with subscript $L$ ($R$) acts to the left (right) of $\rho$, and ${\cal T}$ time reverse-orders operators to the right of $\rho$. Therefore, from (\ref{14}), the density matrix evolution equation is 
\begin{equation}\label{15}
\frac{d\rho(t)}{d t}=-i[H,\rho(t)]-\frac{\lambda}{2}[N,[N,\rho(t)]].
\end{equation}

	With the notation $\overline{\langle A\rangle(t)}\equiv \hbox{Trace}[A\rho(t)]$, we find the coupled equations
\begin{eqnarray}\label{16}
\frac{d\overline{\langle N\rangle(t)}}{d t}&=&ig[\overline{\langle a\rangle(t)}-\overline{\langle a^{\dagger}\rangle(t)}]\\
\frac{d\overline{\langle a\rangle(t)}}{d t}&=&-i\epsilon\overline{\langle a\rangle(t)} -ig -\frac{\lambda}{2}\overline{\langle a\rangle(t)}.  \label{17}
\end{eqnarray}
\noindent Solving  Eqs.(\ref{16}, \ref{17}) and the complex conjugate of (\ref{17}) results in: 
\begin{equation}\label{18}
\overline{\langle N\rangle(t)}=\frac{2g^{2}}{(\lambda/2)^{2}+\epsilon^{2}}\Big\{\frac{\lambda}{2}t-\cos\theta+e^{-\frac{\lambda}{2}t}\cos[\epsilon t-\theta]\Big\}
\end{equation}
\noindent where $\tan\theta=\epsilon/\lambda$.  

	We see from (\ref{18}) that, when $\lambda=0$, there is only oscillation, but when $\lambda\neq0$ there is linear growth of the ensemble mean of the expectation value of the energy $\epsilon N$ and the volume $vN$. The mechanism is that the coupling $\sim g$ causes $N$ to oscillate by a small amount away from its initial value 0, but there is a chance that collapse will occur to an $N\neq 0$ value, and then the oscillation takes it away from that, etc.   (It is worth noting here that "watched pot" behavior is again in evidence,  i.e.,  $\lambda\rightarrow \infty$ implies $\overline{\langle N\rangle}\rightarrow 0$.)
	
	\section {Spacegenerator/Oscillator II}\label{secIV}	
	In the model in the previous section, the mean volume of  space grows linearly with time.  Although inflation models  make space expand exponentially with time,  they do not start from zero space as do  the models under consideration here.  Inflation models assume that enough space had been created earlier so that it may subsequently be described by classical general relativity, with the inflaton field which lives on it obeying quantum field theory.  
	
		There is nothing wrong with an initial space creation that grows linearly with time.  Nonetheless, it is interesting to see that an exponential growth can be obtained with only a slight modification of the model of section \ref{secIII}:  replace the perturbation in the Hamiltonian by $g(a^{\dagger}+a)^{2}$. 

		One then obtains from Eq.(\ref{14}) the coupled equations
\begin{eqnarray}\label{20}
\frac{d\overline{\langle N\rangle}}{d t}&=&2ig[\overline{\langle a^{2}\rangle}-\overline{\langle a^{\dagger 2}\rangle}]\\
\frac{d\overline{\langle a^{2}\rangle}}{d t}&=&-2\overline{\langle a^{ 2}\rangle}[i\epsilon +2ig +2\lambda]
-2ig[1+2\overline{ \langle N\rangle}] \label{21}              
\end{eqnarray}	
\noindent and the complex conjugate of (\ref{21}).  Writing all three quantities as $\sim \exp st$ and putting that into the homogeneous part of the equations results in the cubic equation:
\begin{equation}\label{22}
f(s)\equiv s^{3}+4\lambda s^{2}+(\epsilon^{2}+\lambda^{2}+4\epsilon g)s-32\lambda g^{2}=0.
\end{equation}
		
	For example, if $g=3\lambda/2$ and $\epsilon=[\sqrt{14}-3]\lambda$, Eq.(\ref{22}) becomes 
\[
[s-2\lambda][s-6\lambda e^{ i2\pi/3}][(s-6\lambda e^{- i2\pi/3}]=0.  
\]
	For another example, if $\lambda<<g= \epsilon/4$, then there is a root for which the $s^{3}$ and $s^{2}$ terms may be neglected, with the remaining terms giving $s\approx32\lambda g^{2}/(\epsilon^{2}+4\epsilon g)=\lambda$. 
				
	Generally, it follows from Eq.(\ref{22}) that there are two complex conjugate roots with negative real parts. Because $f(0)=-32\lambda g^{2}<0$ and $f$ has positive slope for $s\geq0$, there is one  solution of (\ref{22}) which is a positive real root, giving rise to an exponential growth of the mean energy and volume of space.  
	
\section{On Choosing the Universe}\label{secX}

	As an indication of how the collapse acts to "choose a universe," in either model, suppose we    neglect the action of the perturbation $\sim g$ over the interval $(t,t_{0})$.  Then, the off-diagonal components of the density matrix decay according to the solution of Eq.(\ref{15}), 
\begin{equation}\label{19}
\langle n|\rho(t)| m\rangle=e^{-(t-t_{0})[i\epsilon(n-m)+\frac{\lambda}{2}(n-m)^{2}]}\langle n|\rho(t_{0}) |m\rangle.  
\end{equation}
\noindent Of course, the density matrix approaching diagonal form does not guarantee that each individual state vector in the ensemble is approaching some definite state $|n\rangle$, but that is the case here\cite{reviews}.

However, one cannot neglect the action of the perturbation: it is working alongside of the collapse. As we have seen in section \ref{2}, when collapse competes with the Hamiltonian evolution, the outcome is not obvious.

What is needed if the collapse is to ``choose a universe?".  A universe is a macroscopic object and, as such, should have well-defined values for macroscopic variables such as the amount of mass it contains. That  is, for any probable state vector in the ensemble, although it is a superposition of states of various $n$-values,  its range of  $n$-values should be small compared to its mean $n$-value. Typically, this requirement is quantified by requiring the deviation from the mean to be small compared to the mean.  

Let's state this requirement in the context of our problem. For any operator $A$, define its expectation value as 
\[
\langle A\rangle(t)\equiv_{w}\negthinspace\negthinspace\langle \Psi,t|A |\Psi, t\rangle_{w}/_{w}\langle \Psi,t|\Psi, t\rangle_{w}. 
\]
Define the ensemble average of products of such expectation values as 
\[
\overline{\langle A\rangle\langle B\rangle...}\equiv\int Dw _{w}\langle \Psi,t|\Psi, t\rangle_{w}\langle A\rangle\langle B\rangle... . 
\]
Then, the squared standard deviation of $N$ is 
\[
(\Delta N)^{2}\equiv\overline{\langle(N-\langle N\rangle)^{2}}\rangle
=\overline{\langle N^{2}\rangle}-\overline{\langle N\rangle^{2}}, 
\]
and the requirement is $\Delta N/\overline{\langle N\rangle}<<1$. 

It should be emphasized that $(\Delta N)^{2}$ is \textit{not}
\[
(\Delta_{0} N)^{2}\equiv\overline{\langle(N-\overline{\langle N\rangle})^{2}}\rangle
=\overline{\langle N^{2}\rangle}-\overline{\langle N\rangle}^{2}.  
\]
Indeed, $(\Delta N)^{2}\leq(\Delta_{0} N)^{2}$, since 
\[
0\leq\overline{[\langle N-\overline{\langle N\rangle}\rangle]^{2}}=\overline{\langle N\rangle^{2}}-\overline{\langle N\rangle}^{2}.
\]

As we have seen, using the density matrix it is easy to calculate ensemble averages involving two state vectors.  However, it is not easy to calculate ensemble averages when four or more state vectors are involved.  

In the two previous sections we have shown how to calculate $\overline{\langle N\rangle}(t)$, and can similarly calculate  $\overline{\langle N^{2}\rangle}(t)$, so  $(\Delta_{0} N)^{2}$ is easy to find. It turns out that, for large $t$,
for both models, $(\Delta_{0} N)^{2}\sim \overline{\langle N\rangle}^{2}$ (i.e., $\sim t^{2}$ for the first and  $\sim \exp2st$ for the second).  

However, $\overline{\langle N\rangle^{2}(t)}$ involves the ensemble average of a product of four state vectors, so $(\Delta N)^{2}$ is not easy to find.  

Here is what we have been able to do.  In Appendix \ref{B}, it is shown in Eq.(\ref{B9}) that
\begin{eqnarray}\label{23}
&&\frac{d}{dt}(\Delta N)^{2}=\frac{d}{dt}(\Delta_{0} N)^{2}\nonumber\\
&&\qquad\qquad +2i\overline{\langle[N-\overline{\langle N\rangle},H]\rangle\langle N-\overline{\langle N\rangle}\rangle}\nonumber\\
&&\qquad\qquad-4\lambda\overline{\langle (N-\langle N\rangle)^{2}\rangle\langle (N-\langle N\rangle)^{2}\rangle}.  
\end{eqnarray}
\noindent This would seem to increase the difficulty since, e.g., the last term in (\ref{23}) is the sum of terms involving the average of products of  four, six and eight  state vectors.  However let us make the not unreasonable assumption that,  for 
most state vectors of high probability, $\langle N-\overline{\langle  N\rangle}\rangle$ is nearly constant, and likewise for  $\langle (N-\langle N\rangle)^{2}\rangle$. Then 
\[
\overline{\langle A\rangle\langle N-\overline{\langle N\rangle}\rangle}\approx \overline{\langle A \rangle}\medspace\overline{\langle N-\overline{\langle N\rangle}\rangle}=0, 
\]
\[
\overline{\langle (N-\langle N\rangle)^{2}\rangle\langle (N-\langle N\rangle)^{2}\rangle}\approx  
[\overline{\langle (N-\langle N\rangle)^{2}\rangle}]^{2},
\]	
\noindent and Eq.(\ref{23}) becomes 
\begin{equation}\label{24}
\frac{d}{dt}(\Delta N)^{2}=\frac{d}{dt}(\Delta_{0} N)^{2}- 4\lambda[(\Delta N)^{2}]^{2}.
\end{equation}
\noindent This is an example of the Riccati equation $dz/dt+z^{2}=f(t)$.  It may be converted to a linear second order equation by the transformation $z=\dot u/u$,  its solution for our two examples for large $t$ involves Bessel functions, etc.  However, the important point is that, for our two examples for large $t$, the left side of Eq.(\ref{24}) is negligible compared to either of the two terms on the right side, so 
\begin{equation}\label{25}
(\Delta N)^{2}\approx \frac{1}{2}\sqrt{\frac{d}{\lambda dt}(\Delta_{0} N)^{2}}.
\end{equation}

	For the example in section \ref{secIII}, we may write, $(\Delta_{0} N)^{2}\approx (N_{0}t/T_{0})^{2}$ for large $t$, where $N_{0}$ is the current number of Planck masses in the universe and $T_{0}$ is the current age of the universe.  We therefore find from Eq.(\ref{25}) that 
\[
(\Delta N)^{2}\approx\frac{N_{0}}{\sqrt{2\lambda T_{0}}}\thickspace
\hbox{and\medspace} \frac{\Delta N}{\overline{\langle N\rangle}}\approx\frac{1}{[2\lambda T_{0}]^{1/4}N_{0}^{1/2}}. 
\]
	
For the example in section \ref{secIV}, we may write, $(\Delta_{0} N)^{2}\approx [N_{0}\exp s_{0}(t-T_{0})]^{2}$ for large $t$, where $s_{0}N_{0}$ is the current rate of increase of Planck masses (see the next section for more discussion of this).  We therefore find from Eq.(\ref{25}) that 		
\[
(\Delta N)^{2}\approx\frac{N_{0}}{\sqrt{2\lambda/ s_{0}}}\thickspace
\hbox{and\medspace} \frac{\Delta N}{\overline{\langle N\rangle}}\approx\frac{1}{[2\lambda /s_{0}]^{1/4}N_{0}^{1/2}}. 
\]	

As mentioned in the next section, it is plausible that $\lambda T_{0}$ and $\lambda /s_{0}$ are of order 1.  Therefore, indeed, $\Delta N/\overline{\langle N\rangle}\approx N_{0}^{-1/2}<<1$.  
	
\section{Concluding Remarks}\label{secV}

	By means of a simple quantum mechanical model, the usefulness of collapse dynamics in creating the universe out of nothing after a long wait, generating its energy, and choosing a macroscopic universe out of a superposition, has been sketched.  The model may be regarded as just an illustration, but one might try to take it more seriously.  

	It is perhaps natural to choose  $\lambda^{-1}$ and $g^{-1}$ to be of the order of the Planck time $T_{p}$.  If one takes the volume of the present universe to be $V_{0}\approx 3\times 10^{80}$ cubic meters, takes the mass-energy density to be the present critical density value  $\rho_{0}\approx 10^{-26}$kg/m$^{3}$ then, with the Planck mass $\epsilon=M_{p}\approx 2\times 10^{-8}$kg, there are presently $N_{0}=\rho_{0}V_{0}/M_{p}\approx 10^{62}$ Planck masses  in the universe.  
	
	Suppose that all this mass-energy is generated at the beginning of the universe, according to the model of section \ref{secIV}, $N(t)\approx\exp(t/T_{p})$.  It would take  time $t\approx 140T_{p}$ to generate $N_{0}$ such masses. If each Planck mass occupies a Planck volume, the universe at that time would be of  radius $\approx (3N_{0}/4\pi)^{1/3}L_{p}\approx 5\times10^{-13}$cm  ($L_{p}\approx 1.6\times 10^{-33}$cm is the Planck length), about 5 times larger than the radius of a proton.  Of course, with this scenario, one wants the process to abruptly stop at this time. One might make $g$, $\lambda$ and $\epsilon$ proportional to $\Theta(N_{0}-N)$, and let a different Hamiltonian take over the subsequent evolution but this seems rather artificial. 
	
	More interesting would be if the process were ongoing, with $\lambda$ a  suitably decreasing function of $N$, to provide a mechanism which continually slows the energy production down, yielding the correct amount  of energy and energy production rate at the present era. 
 
 With regard to the current energy production rate, here is a suggestive result. 
 
 According to the model, given the present time $T_{0}$, the number of Planck masses at time $t\geq T_{0}$  is $N(t)= N_{0}\exp \lambda_{0}(t-T_{0})$, where $\lambda_{0}$ is the current collapse rate.  Thus, since the energy in the universe is $E(t)=M_{p}N(t)$, its present rate of change is  
 \[
 dE/dt=E_{0}\lambda_{0}.
 \]
On the other hand, suppose next that the radius of the universe is expanding with the speed of light, $dR/dt=c$. Then, the present rate of increase of energy in the universe is 
 \[
dE/dt=\rho_{0}dV/dt=\rho_{0}4\pi R_{0}^{2}c=3E_{0}c/R_{0}.  
 \]  
 \noindent Equating these two expressions for $dE/dt$, using $R_{0}\approx (3V_{0}/4\pi)^{1/3}=4\times 10^{26}$m, we obtain 
\[ 
 \lambda_{0}=3c/R_{0}\approx 2\times 10^{-18}\hbox{sec}^{-1},  
\] 
approximately the inverse of the age of the universe.  

	This is not far from the value  $\lambda\approx 10^{-16}$sec$^{-1}$ suggested for the GRW theory\cite{GRW} and adopted for CSL.  It suggests  considering the consequences if the collapse rate in CSL, hitherto considered a constant is, instead, the inverse of the age of the universe.  This would imply a large rate of excitation of matter at the beginning of the universe.  However, it would not have much effect, either on visible excitation or on the collapse of  macroscopic objects  over a time interval, say, 1/100 to 100 times the present age of the universe.  

The reader will have noted that there has been no mention of general relativity.  Connection with the FRW formalism can be made through identifying an operator for the ``radius R" of the universe.    

$R$ depends not only on the operator $N(t)$ but also on the operator representing the volume $v(t)$ of a cell.  $v$ must have dynamics because, if it equals the Planck volume at the beginning of the universe, it does not remain so.  The current value of $v$ may be found as follows.  The change of $V$ is $dV=vdN$, while the change of energy is $dE=\epsilon dN$.  Since $dE/dV=\rho_{0}$, we obtain $v=\epsilon/ \rho_{0}\approx 2\times 10^{18}$m$^{3}$, i.e., cells of Planck energy which are currently added have length $\approx 1000$km.  

So, there must be an addition to the Hamiltonian to give dynamics for the operator $v$, presumably guided by the FRW formalism, with 
\[
\frac{4\pi}{3}R^{3}(t)=\int_{0}^{t}dt'\frac{dN(t')}{dt'}v(t').
\]
Moreover, if the created energy is to include not only dark energy but also visible matter and dark matter, dynamics must be included describing how the latter two evolve from the total energy $N\epsilon$.  	
 	
\acknowledgements

I would like to thank Fay Dowker for a stimulating e-mail correspondence six years ago which resulted in   the analysis presented in section \ref{sec2} and Appendix \ref{A}. 	
	
	\appendix

\section{Spin Problem}\label{A}
	Consider an initially normalized state vector $|\Psi, t\rangle$ describing a spin in the $x-z$ plane. If $0\leq\theta\leq 2\pi$ is the angle between the $z-$axis and the spin vector, the components of the spin in the $\sigma_{z}$ basis are $\langle \uparrow|\Psi, t\rangle=\cos(\theta/2)$,  $\langle \downarrow|\Psi, t\rangle=\sin(\theta/2)$ (up to a phase factor). The spin part of the evolution given in Eq. (\ref{1}), for an infinitesimal time interval $dt$, is:
\begin{equation}\label{A1}
|\Psi, t+dt\rangle_{w}=e^{-dt\{\frac{i}{2}\omega\sigma_{2} +\frac{1}{4\lambda}[w(t)-2\lambda\sigma_{1}]^{2}\}}|\Psi,t\rangle_{w}.   
\end{equation}
\noindent Eq. (\ref{A1}), expressed in the $\sigma_{z}$ basis, retaining all terms necessary for first order in $dt$ is: 
\begin{eqnarray}\label{A2}
&&A\cos[\theta(t+dt)/2]=e^{-\frac{1}{4\lambda}w^{2}(t)}\{\cos(\theta/2)[1+w(t)dt \nonumber\\
&&\qquad\qquad+\frac{1}{2}(w(t)dt)^{2}
-\lambda dt]-\sin(\theta/2)[\frac{1}{2}\omega dt]\}  \nonumber \\
&&A\sin[\theta(t+dt)/2]=e^{-\frac{1}{4\lambda}w^{2}(t)}\{\sin(\theta/2)[1-w(t)dt \nonumber\\
&&\qquad\qquad+\frac{1}{2}(w(t)dt)^{2}
-\lambda dt]+\cos(\theta/2)[\frac{1}{2}\omega dt]\}.  
\end{eqnarray}
\noindent The amplitude $A$ is necessary because the state vector evolution is not unitary.  Indeed, this is what makes the probability rule Eq. (\ref{2}) meaningful: 
 \begin{eqnarray}\label{A3}
&&P(w(t))\sim_{w}\negthinspace\negthinspace
\langle\Psi,t|\Psi, t\rangle_{w}=A^{2}\nonumber\\
&&=e^{-\frac{1}{2\lambda}w^{2}(t)}\{1+2w(t)dt[\cos^{2}(\theta/2)-\sin^{2}(\theta/2)]\nonumber\\
&&+2[\frac{1}{2}(w(t)dt)^{2}
-\lambda dt]+(w(t)dt)^{2}\}\nonumber\\
&&=e^{-\frac{1}{2\lambda}w^{2}(t)}[1+2w(t)dt\cos \theta]
\end{eqnarray}
\noindent We note that the total probability is 1, as may be seen by multiplying Eq. (\ref{A3}) by $(2\pi\lambda/dt)^{-1/2}dw(t)$ and integrating over $(-\infty,\infty)$.   In the last step of Eq. (\ref{A3}), we have set $(w(t)dt)^{2}=\lambda dt$, since that is what its integral gives to order $dt$ in all of our calculations (the same thing could have been done in Eq. (\ref{A2}) and shall be done in subsequent equations).  

	Solving Eqs. (\ref{A2}) to order $dt$ yields: 
\begin{eqnarray}\label{A4}
\theta(t+dt)&=&2\tan^{-1}\{\tan(\theta/2)+\frac{1}{2}\omega dt\sec^{2}(\theta/2)\nonumber\\
&+&2[\lambda dt-w(t)dt]\tan(\theta/2)\}
\end{eqnarray}
 \noindent Using $\tan^{-1}(\tan\phi+x)=\phi+x\cos^{2}\phi-x^{2}\sin\phi\cos^{3}\phi+...$ in Eq. (\ref{A4}) results in:
 \begin{equation}\label{A5}
d\theta(t)=\omega dt-2w(t)dt\sin\theta+2\lambda dt\sin\theta \cos\theta
\end{equation}

	Now that we have Eqs.(\ref{A3}, \ref{A5}), we can find $\overline{d\theta}$, $\overline{(d\theta)^{2}}$, which are used in constructing the Fokker-Planck equation for the probability distribution of $\theta$ in section \ref{sec2}: 
\begin{eqnarray}\label{A6}
\overline{d\theta}&\equiv&\int_{-\infty}^{\infty}\frac{1}{\sqrt{2\pi\lambda/dt}}dw(t)P(w(t))d\theta(t)\nonumber\\
&&\qquad\qquad\qquad=\omega dt-2\lambda dt\sin\theta \cos\theta\\
\overline{(d\theta)^{2}}&\equiv&\int_{-\infty}^{\infty}\frac{1}{\sqrt{2\pi\lambda/dt}}dwP(w(t))(d\theta)^{2}(t)\nonumber\\
&&\qquad\qquad\qquad=4\lambda dt \sin^{2}\theta.\label{A7}
\end{eqnarray}

\section{Solution of Eq.(\ref{4}) for $\omega=0$}\label{C}
For convenience, we give here the solution\cite{pearle76} of 
\begin{equation}\label{C1}
\frac{\partial}{\partial t}R=2\lambda\Big\{\frac{\partial}{\partial \theta}[\sin\theta\cos\theta]R+\frac{\partial^{2}}{\partial \theta^{2}}\sin^{2}\theta R\Big\}. 
\end{equation}

	By changing variables to the squared amplitude $x=\cos^{2}(\theta/2)$, and $R=(1/2)\sin\theta R'$, the equivalent equation is obtained:
\begin{equation}\label{C2}
\frac{\partial}{\partial t}R'=8\lambda\frac{\partial^{2}}{\partial x^{2}}[x(1-x)]^{2} R'. 
\end{equation}
\noindent where $0\leq x\leq 1$.  Eq.(\ref{C2}) clearly shows the no-drift condition $d\overline{x}(t)/dt=0$ 
necessary for Born-rule-obeying collapse, as discussed in section \ref{subIIA}.  It also shows how the diffusion vanishes at $x=0, 1$, providing absorbing boundary conditions.  

A further change of variables, $z=\ln[x/(1-x)]$ and $R'=x(1-x)R''$ provides an equation with normal diffusion,
\begin{equation}\label{C3}
\frac{\partial}{\partial t}R''=8\lambda\Big\{\frac{\partial^{2}}{\partial z^{2}}R''-\frac{\partial}{\partial z}
\tanh(z/2)R''\Big\},  
\end{equation}
\noindent which is readily solved:
\begin{eqnarray}\label{C4}
&&R''(z)=\frac{1}{8\sqrt{2\pi\lambda t}\cosh(z_{0}/2)}\Big\{e^{z_{0}/2}e^{\frac{-1}{32\lambda t}[z-z_{0}-8\lambda t]^{2}}\nonumber\\
&&\qquad\qquad\qquad\qquad+e^{-z_{0}/2}e^{\frac{-1}{32\lambda t}[z-z_{0}+8\lambda t]^{2}}\Big\}. 
\end{eqnarray}
We see from Eq.(\ref{C4}) that $R''(0)\rightarrow\delta(z-z_{0})$ and 
\[
R''(\infty )\rightarrow\frac{1}{\cosh(z_{0}/2)}\Big\{e^{z_{0}/2}\delta(z-\infty)+e^{-z_{0}/2}\delta(z+\infty)\Big\}.
\]
\noindent In terms of $\theta$, the solution (\ref{C4}) of Eq.(\ref{C1}) is
\begin{eqnarray}\label{C5}
&&R(\theta)d\theta=\frac{1}{\sqrt{8\pi\lambda}}d \ln\tan(\theta/2)\nonumber\\
&&\Big\{\cos^{2}(\theta_{0}/2)e^{\frac{-1}{8\lambda t}\big[\ln\frac{\tan(\theta/2)}{\tan(\theta_{0}/2}
+4\lambda t\big]^{2}}\nonumber\\
&&\qquad\qquad+\sin^{2}(\theta_{0}/2)e^{\frac{-1}{8\lambda t}\big[\ln\frac{\tan(\theta/2)}{\tan(\theta_{0}/2}-4\lambda t\big]^{2}}\Big\}.
\end{eqnarray}
\noindent  We see from Eq.(\ref{C5}) that $R(0)\rightarrow\delta(\theta-\theta_{0})$ and note the collapse behavior obeying the Born rule,
\[
R''(\infty )\rightarrow\cos^{2}(\theta_{0}/2)\delta(\theta)+\sin^{2}(\theta_{0}/2)\delta(\theta-\pi).
\]

\section{Ensemble Averages of Products}\label{B}

We first prove the following theorem.  Consider the state vector evolution (\ref{13}).  Then 
\begin{eqnarray}\label{B1}
&&\frac{d}{dt}\overline{\langle A\rangle\langle B\rangle}=
\overline{\langle-i[A,H]-(\lambda/2)[N,[N,A]]\rangle\langle B\rangle}\nonumber\\
&&\qquad+\overline{\langle A\rangle\langle-i[B,H]-(\lambda/2)[N,[N,B]]\rangle}\nonumber\\
&&+4\lambda\overline{\langle N\rangle^{2}\langle A\rangle\langle B\rangle}\nonumber\\
&&-2\lambda[\overline{\langle N\rangle\langle A\rangle\langle BN+ NB\rangle}+\overline{\langle N\rangle\langle B\rangle\langle AN+ NA\rangle}]\nonumber\\
&&+\lambda[\overline{\langle AN+ NA\rangle\langle BN+ NB\rangle}].
\end{eqnarray}
\noindent In Eq.(\ref{B1}), $\langle Q\rangle(t)\equiv _{w}\negthinspace\negthinspace\langle \Psi,t|Q |\Psi, t\rangle_{w}/_{w}\langle \Psi,t|\Psi, t\rangle_{w}$. The overline refers to the ensemble average:   for example,   $\overline{\langle A\rangle}(t)=\int Dw _{w}\langle \Psi,t|\Psi, t\rangle_{w}\langle A\rangle(t)\equiv$Tr$A\rho(t)$, where $\rho(t)$ is the density matrix.

	According to Eq.(\ref{13}), the state vector at time $t+dt$ is
 \begin{eqnarray}\label{B2}
&&|\Psi, t+dt\rangle_{w',w}=e^{-idtH-\frac{dt}{4\lambda} [w'-2\lambda N]^{2}}|\Psi, t\rangle_{w}=e^{-\frac{1}{4\lambda}dtw'^{2}}\nonumber\\
&&\qquad\qquad\cdot[1-idtH +dtw'N-(\lambda/2)dtN^{2}]|\Psi, t\rangle_{w}.
\end{eqnarray}
\noindent In Eq.(\ref{B2}), we have written $w'$ as the scalar field at time $t+dt$, and $w$ represents the scalar field for $0\leq t\leq t$.  We have also set $(w'dtN)^{2}/2=\lambda dtN^{2}/2$,  and will continue to do so, since that is what it will equal when integrated over the probability. 

	The ensemble average of the expectation value of $A$  at time $t+dt$ (calculated using Eq.(\ref{B2})),  is obtained by multiplying  the expectation value, 
\begin{eqnarray}\label{B3}
&&\langle  A\rangle(t+dt)=\langle  A\rangle(t) +dt\{-i\langle[A,H]\rangle(t)+w'\langle\{N,A\}_{+}\rangle(t)\nonumber\\
&&\qquad-(\lambda/2)\langle[N,[N,A(t)]]\rangle(t)\}\nonumber\\
&&\qquad\qquad\cdot\Big\{\frac{\medspace _{w}\langle\Psi, t |\Psi, t\rangle_{w}}{\medspace _{w}\langle\Psi, t |(1+2dtw'N)|\Psi, t\rangle_{w}}\Big\}
\end{eqnarray}
\noindent by the probability density 
\begin{eqnarray}\label{B4}
&&\thinspace_{w',w}\langle \Psi,t+dt|\Psi, t+dt\rangle_{w',w}=
e^{-\frac{1}{2\lambda}dtw'^{2}}\nonumber\\
&&\quad\qquad\qquad\qquad\cdot _{w}\langle\Psi, t |e^{2dtw'N-2\lambda N^{2}}|\Psi, t\rangle_{w}\nonumber\\
&&\qquad= e^{-\frac{1}{2\lambda}dtw'^{2}}\medspace _{w}\langle\Psi, t |(1+2dtw'N)|\Psi, t\rangle_{w}
\end{eqnarray}
\noindent and by $dw'Dw$, and integrating over all $w',w$, to obtain the well-known result:
\begin{eqnarray}\label{B5}
\frac{d}{dt}\overline{\langle  A\rangle(t)}=\hbox{Tr}\Big[\{-i[A,H]-(\lambda/2)[N,[N,A(t)]]\}\rho(t)\Big].  
\end{eqnarray}
 
Similarly, the ensemble average of the product of two density matrices at time $t+dt$ is
\begin{eqnarray}\label{B6}
&&\overline{\langle  A\rangle(t+dt)\langle  B\rangle(t+dt)}=\int dw'Dwe^{-\frac{1}{2\lambda}dtw'^{2}}
_{w}\langle\Psi, t |\Psi, t\rangle_{w}\nonumber\\
&&\cdot \Big\{\langle  A\rangle(t) +dt\{-i\langle[A,H]\rangle(t)+w'\langle\{N,A\}_{+}\rangle(t)\nonumber\\
&&\qquad-(\lambda/2)\langle[N,[N,A(t)]]\rangle(t)\}\Big\}\nonumber\\
&&\cdot\Big\{\langle  B\rangle(t) +dt\{-i\langle[B,H]\rangle(t)+w'\langle\{N,B\}_{+}\rangle(t)\nonumber\\
&&\qquad-(\lambda/2)\langle[N,[N,B(t)]]\rangle(t)\}\Big\}\nonumber\\
&&\qquad\qquad\cdot\Big\{\frac{\medspace _{w}\langle\Psi, t |\Psi, t\rangle_{w}}{\medspace _{w}\langle\Psi, t |(1-2dtw'N)|\Psi, t\rangle_{w}}\Big\}. 
\end{eqnarray}
\noindent Writing the last bracket in Eq.(\ref{B6}) as
\begin{eqnarray}\label{B7}
\Big\{\frac{\medspace _{w}\langle\Psi, t |\Psi, t\rangle_{w}}{\medspace _{w}\langle\Psi, t |(1-2dtw'N)|\Psi, t\rangle_{w}}\Big\}&=&1+2dtw'\langle  N\rangle(t)\nonumber\\
&&+4\lambda dt\langle  N\rangle^{2}(t), 
\end{eqnarray}
\noindent the integrals can be performed with the result Eq.(\ref{B1}).

	One may check that, if $B=1$ is inserted into Eq.(\ref{B1}), then Eq.(\ref{B5}) is obtained.  

	Now, set $A=B=N$.  Eq.(\ref{B1}) becomes: 
\begin{eqnarray}\label{B8}
&&\frac{d}{dt}\overline{\langle N\rangle^{2}(t)}=
-2i\overline{\langle[N,H]\rangle\langle N\rangle}\nonumber\\
&&\qquad+4\lambda\overline{\langle (N-\langle N\rangle)^{2}\rangle\langle (N-\langle N\rangle)^{2}\rangle}.  
\end{eqnarray}
\noindent Since 
\[
\overline{\langle(N-\langle N\rangle)^{2}\rangle}=\overline{\langle N^{2}\rangle}-\overline{\langle N\rangle^{2}}
\]
\noindent and
\[
\overline{\langle(N-\overline{\langle N\rangle})^{2}\rangle}=\overline{\langle N^{2}\rangle}-\overline{\langle N\rangle}^{2}, 
\]
\noindent it follows from Eq.(\ref{B8}) that 
\begin{eqnarray}\label{B9}
&&\frac{d}{dt}\overline{\langle(N-\langle N\rangle)^{2}\rangle}=\frac{d}{dt}\overline{\langle(N-\overline{\langle N\rangle})^{2}\rangle}\nonumber\\
&&\qquad\qquad +2i\overline{\langle[N-\overline{\langle N\rangle},H]\rangle\langle (N-\overline{\langle N\rangle}\rangle}\nonumber\\
&&\qquad\qquad-4\lambda\overline{\langle (N-\langle N\rangle)^{2}\rangle\langle (N-\langle N\rangle)^{2}\rangle}.  
\end{eqnarray}
\noindent which is used in section V, Eq.(\ref{23}).


\begin{thebibliography}{99}

\bibitem{CSL} P. Pearle, Phys. Rev. A{\bf 39}, 2277 (1989):  G. C. Ghirardi, P. Pearle and A. Rimini,  Phys. Rev. A{\bf 42}, 78 (1990).

\bibitem{reviews} Some reviews are: P. Pearle in \textit{Open systems and measurement in
relativistic quantum theory}, F. Petruccione and H. P. Breuer eds., p.195 (Springer Verlag, Heidelberg 1999):   A. Bassi and G. C. Ghirardi, \textit{Physics Reports} {\bf379}, 257 (2003):  P. Pearle, Journ. Phys. A: Math Theor. 40, 3189 (2007) and continued in \textit{Quantum Reality, Relativistic Causality and Closing the Epistemic Circle}, eds. W. Myrvold and J. Christian (Springer, 2009), pp.257-292. 

\bibitem{pearlesquiresetc.}  It is a remarkable property of CSL that, because the collapse rate of particles of mass $m$ is governed by their mass density operator (as opposed to the number density operator multiplied by something other than $m$), the energy production rate is quite small: see  P. Pearle and E. Squires, Phys. Rev. Lett {\bf73}, 1 (1994).  Verification/refutation of CSL may come about due to future highly sensitive experiments which detect/do not detect such an energy increase. 

\bibitem{IB} A.  Bassi and E. Ippoliti,  Phys. Rev. {\bf A69}, 012105 (2004)

\bibitem{pearle76} P. Pearle, Phys. Rev. {\bf D13}, 857 (1976).

\bibitem{Pearletimetoreduce} P. Pearle, Journ. of Stat. Phys. {\bf 41}, 719 (1985).



\bibitem{GRW} G. C. Ghirardi, A. Rimini and T. Weber, Phys. Rev. D{\bf 34}, 470 (1986). 

\end{thebibliography}
\end{document}